\documentclass[a4paper,10pt]{article}
\usepackage[cp1251]{inputenc}
\usepackage[english]{babel}
\usepackage{amsmath}
\usepackage{amssymb}
\usepackage{graphicx}
\usepackage[pdftex]{color}
\usepackage[square,numbers,comma]{natbib}
\usepackage[pdftex]{hyperref}
\hypersetup{
             final,
             colorlinks=true,
             linkcolor=blue,
             citecolor=red
}

\begin{document}
\twocolumn[\scriptsize{\slshape ISSN 0021-3640, JETP Letters, 2008,
Vol. 87, No. 6, pp. 292–298. \textcopyright\, Pleiades Publishing,
Ltd., 2008.}

\scriptsize{\slshape Original Russian Text \textcopyright\, P.V.
Ratnikov, 2008, published in Pis’ma v Zhurnal
\'{E}ksperimental’no\u{\i} i Teoretichesko\u{\i} Fiziki, 2008, Vol.
87, No. 6, pp. 343–349.}

\vspace{0.5cm}

\hrule\vspace{0.07cm}

\hrule

\vspace{0.75cm}

\begin{center}
\LARGE{\bf Transition of Graphene on a Substrate}

\LARGE{\bf to a Semimetallic State}

\vspace{0.25cm}

\large{\bf P. V. Ratnikov}

\vspace{0.1cm}

\normalsize

\textit{Lebedev Physical Institute, Russian Academy of Sciences,}

\textit{Leninski\u{\i} pr. 53, Moscow, 119991 Russia}

\textit{e-mail: ratnikov@lpi.ru}

Received December 13, 2007; in final form, February 1, 2008
\end{center}

\vspace{0.1cm}
\begin{list}{}
{\rightmargin=1cm \leftmargin=1cm}
\item
\small{Graphene on a substrate has been shown to exhibit a
transition, depending on the substrate material, from a zero-gap
semiconductor state to a semimetallic state. The ground-state energy
of the electron (hole) gas has been calculated within the
random-phase approximation.}

\vspace{0.05cm}


\vspace{0.05cm}

\small{\bf DOI}: 10.1134/S0021364008060064

\end{list}\vspace{0.5cm}]

\begin{center}
1. INTRODUCTION
\end{center}

A monolayer of carbon atoms that form a perfect hexagonal lattice
(graphene) has such a band structure that the energy gap is equal to
zero in three $K$ and three $K^\prime$ points of the Brillouin zone
\citep{Novoselov}. The Schr\"{o}dinger equation in the ${\bf
k}\cdot{\bf p}$ approximation has the form \citep{Ando}
\begin{equation}\label{1}
\widehat{H}_0F({\bf r})=\varepsilon F({\bf r}),
\end{equation}
\begin{equation}\label{2}
\widehat{H}_0=\begin{pmatrix} u{\boldsymbol\sigma}\cdot\widehat{\bf
k}&0\\
0&u{\boldsymbol\sigma}\cdot\widehat{\bf k}^\prime
\end{pmatrix},
\end{equation}
where the quantity $u=\frac{3}{2}\gamma a_0\approx9.84\times10^7$\,
cm/s is similar to the Kane matrix element for the rate of interband
transitions in the Dirac model \citep{Volkov}, $\gamma\simeq3$ eV is
the band parameter numerically equal to the overlap integral of
atomic orbitals that participate in the chemical bonds of carbon
atoms in graphene, $a_0=1.44$\,$\text{\AA}$ is the interatomic
distance in the graphene lattice, ${\boldsymbol\sigma}=(\sigma_1,
\sigma_2)$ are the Pauli matrices, $\widehat{{\bf
k}}=(\widehat{k}_x, \widehat{k}_y)$, $\widehat{{\bf
k}}^\prime=(\widehat{k}_x, -\widehat{k}_y), \
\widehat{k}_{x,y}=-i\partial_{x,y},$ and $\hbar=1$.

The unitary transformation $\widehat{U}_1=\bigl(\begin{smallmatrix}
I & 0\\0 & \sigma_2
\end{smallmatrix}\bigr)$, where $I$ is the $2\times2$ unit matrix,
reduces the Hamiltonian $\widehat{H}_0$ to the form
\begin{equation}\label{3}
\widehat{H}^\prime_0=\widehat{U}_1\widehat{H}_0\widehat{U}^\dagger_1=
\begin{pmatrix}
u{\boldsymbol\sigma}\cdot\widehat{\bf
k}&0\\
0&-u{\boldsymbol\sigma}\cdot\widehat{\bf k}
\end{pmatrix},
\end{equation}
The corresponding equation with the transformed wavefunction
$\varphi({\bf r})=\widehat{U}_1F({\bf r})$ is equivalent to a pair
of Weyl equations\footnote{The first application of the Weyl
equation to the description of the charge carriers in a zero-gap
semiconductor was probably proposed in \citep{Tsvelik} in view of
the appearance of the axial Adler–Bell–Jackiw anomaly in parallel
electric and magnetic fields.}. In quantum electrodynamics (QED),
the Weyl equation describes \textsl{neutrino}, a massless spin-1/2
particle. Since the particles were assumed to be spinless when
deriving Eq. \eqref{1}, Novoselov \citep{Novoselov} introduced the
notion of \textsl{pseudospin}. As correctly mentioned in
\citep{Dharma-wardana}, a complete model should include an
$8\times8$ matrix Hamiltonian due to the twofold valley degeneracy,
twofold pseudospin degeneracy, and twofold spin degeneracy. The
$8\times8$ Hamiltonian may be reduced to the $4\times4$ Hamiltonian
given by Eq. \eqref{3} if the Fermi momentum is determined in the
spin-unpolarized state of the particles. In this case, the
degeneracy multiplicity is $\nu=\nu_{e,h}=2$ ($\nu_{e,h}$ is the
valley degeneracy of the conduction band or the valence band).

To obtain the Dirac equation, let us perform another unitary
transformation
$\widehat{U}_2=\frac{1}{\sqrt{2}}\bigl(\begin{smallmatrix} I & I\\I
& -I
\end{smallmatrix}\bigr)$
\begin{equation}
\widehat{H}^{\prime\prime}_0=\widehat{U}_2\widehat{H}^\prime_0\widehat{U}^\dagger_2=
\begin{pmatrix}
0&u{\boldsymbol\sigma}\cdot\widehat{\bf k}\\
u{\boldsymbol\sigma}\cdot\widehat{\bf k}&0
\end{pmatrix}\equiv u{\boldsymbol\alpha}\cdot\widehat{\bf k},
\end{equation}
where ${\boldsymbol\alpha}=\bigl(\begin{smallmatrix}0&{\boldsymbol\sigma}\\
{\boldsymbol\sigma}&0\end{smallmatrix}\bigr)$ are the Dirac
matrices. Thus, the charge carriers in graphene are described in the
framework of the zero-gap ($\Delta=0$) Dirac model. The use of the
Dirac equation as a $4\times4$ matrix equation in the twodimensional
system is justified because the $4\times4$ and $2\times2$ matrix
representations may be equally used in the case of two spatial
dimensions \citep{Tsvelik}. This allows us to apply the diagram
technique of QED to the case of the two-dimensional system of Dirac
fermions (graphene).

In this work, it is shown that graphene on a substrate may undergo a
transition from a zero-gap semiconductor state to a semimetallic
state, depending on the substrate material. The transition occurs at
$\alpha^*\simeq1$, where $\alpha^*=\frac{e^2}{\kappa_{eff}u}$ is the
analog of the fine-structure constant, whose numerical value depends
on the relative permittivity $\epsilon_{1,2}$  of the media
surrounding the graphene and
$\kappa_{eff}=\frac{\epsilon_1+\epsilon_2}{2}$, similar to a thin
film \citep{Keldysh1}.

\newpage

\begin{center}
2. GROUND-STATE ENERGY\\ (GENERAL CONSIDERATION)
\end{center}

To determine whether the zero-gap semiconductor phase of graphene is
stable with respect to the transition to another phase, one has to
calculate the ground-state energy of the electron (hole) gas that
appears in graphene when the electric field is applied. The
ground-state energy per particle is the sum of the three terms
\begin{equation}
E_{gs}=E_{kin}+E_{exch}+E_{corr}.
\end{equation}
Here, the average kinetic energy is given by the expression
$E_{kin}=\frac{2}{3}up_F$ because, according to Eq. \eqref{1}, the
dispersion relation of the charge carriers is linear near the $K$
and $K^\prime$ points of the Brillouin zone: $\varepsilon_{\bf
p}=\pm u|{\bf p}|$ ($+$ and $-$ correspond to electrons and holes,
respectively), $p_F=\sqrt{2\pi n_{2D}/\nu}$ is the Fermi momentum,
$n_{2D}$ is the areal density of particles, and $\nu$ is the
above-mentioned degeneracy multiplicity. If the Fermi level
$\varepsilon_F$ lies above $\varepsilon=0$, the charge carriers in
the system are only the conduction electrons with a number of
valleys $\nu_e=2$; if $\varepsilon_F<0$, then the charge carriers
are only holes with $\nu_h=2$. The position of the Fermi level may
be changed by applying the electric field \citep{Novoselov}. Both
cases are obviously equivalent in the Dirac model. Below, we will
consider the case of electrons for definiteness.

The exchange energy is given by the diagram (see
\hyperlink{fig1}{Fig. 1})
\begin{equation*}
E_{exch}=-\frac{\nu}{2n_{2D}}\int\frac{d^2{\bf
p}d\varepsilon}{(2\pi)^3} \frac{d^2{\bf
k}d\omega}{(2\pi)^3}Sp\left\{\Gamma^\mu ({\bf p}, \varepsilon; {\bf
k}, \omega)\right.
\end{equation*}
\begin{equation}\label{6}
\left.\times G\left({\bf p}, \varepsilon\right)\gamma^\nu
G\left({\bf k}, \omega\right)\right\}D^{(0)}_{\mu\nu}\left({\bf
p}-{\bf k}, \varepsilon-\omega\right),
\end{equation}
where the photon propagator is $D^{(0)}_{\mu\nu}\left({\bf p}-{\bf
k}, \varepsilon-\omega\right)\approx V\left({\bf p}-{\bf
k}\right)\delta_{\mu4}\delta_{\nu4}$ (we neglect the photon poles
whose contributions to the integral with respect to the frequencies
$\varepsilon$ and $\omega$ are on the order of
$\left(u/c\right)^2\sim10^{-5}$ each, i.e., small compared to the
contribution of the Green’s function poles), $V\left({\bf
q}\right)=\frac{2\pi e^2}{\kappa_{eff}|{\bf q}|}$ is the Coulomb law
in the two-dimensional case.

The correlation energy is given by the formula\footnote{Formula
\eqref{7} is derived in the nonrelativistic case and is the sum of
the ring diagrams of all orders (as the most divergent diagrams).
The situation in the relativistic case is similar; therefore, the
same formula with the corresponding polarization operators is used
here.} \citep{Keldysh}
\begin{equation*}\label{7}
E_{corr}=\frac{1}{2n_{2D}}\int\frac{d^2{\bf k}d\omega}{(2\pi)^3}
\end{equation*}
\begin{equation}\label{7}
\times\int\limits_0^1\frac{d\lambda}{\lambda}
\left[\frac{-\lambda\nu V({\bf k})\Pi_{44}({\bf k},
i\omega)}{1-\lambda\nu V({\bf k})\Pi_{44}({\bf k},
i\omega)}+\lambda\nu V({\bf k})\Pi_{44}({\bf k}, i\omega)\right].
\end{equation}

The polarization operator $\Pi^{(0)}_{44}({\bf k}, i\omega)$ in the
lowest order in the interaction (see \hyperlink{fig2}{Fig. 2}) is
calculated from the zero-approximation Green’s functions
\citep{Markova}
\begin{equation}\label{8}
G^{(0)}({\bf p}, \varepsilon)=-\frac{u\widehat{p}}{(\varepsilon_{\bf
p}-\varepsilon+i\delta_{-})(\varepsilon_{\bf
p}+\varepsilon-i\delta_{+})},
\end{equation}
\begin{center}
\hypertarget{fig1}{}
\includegraphics[scale=0.12]{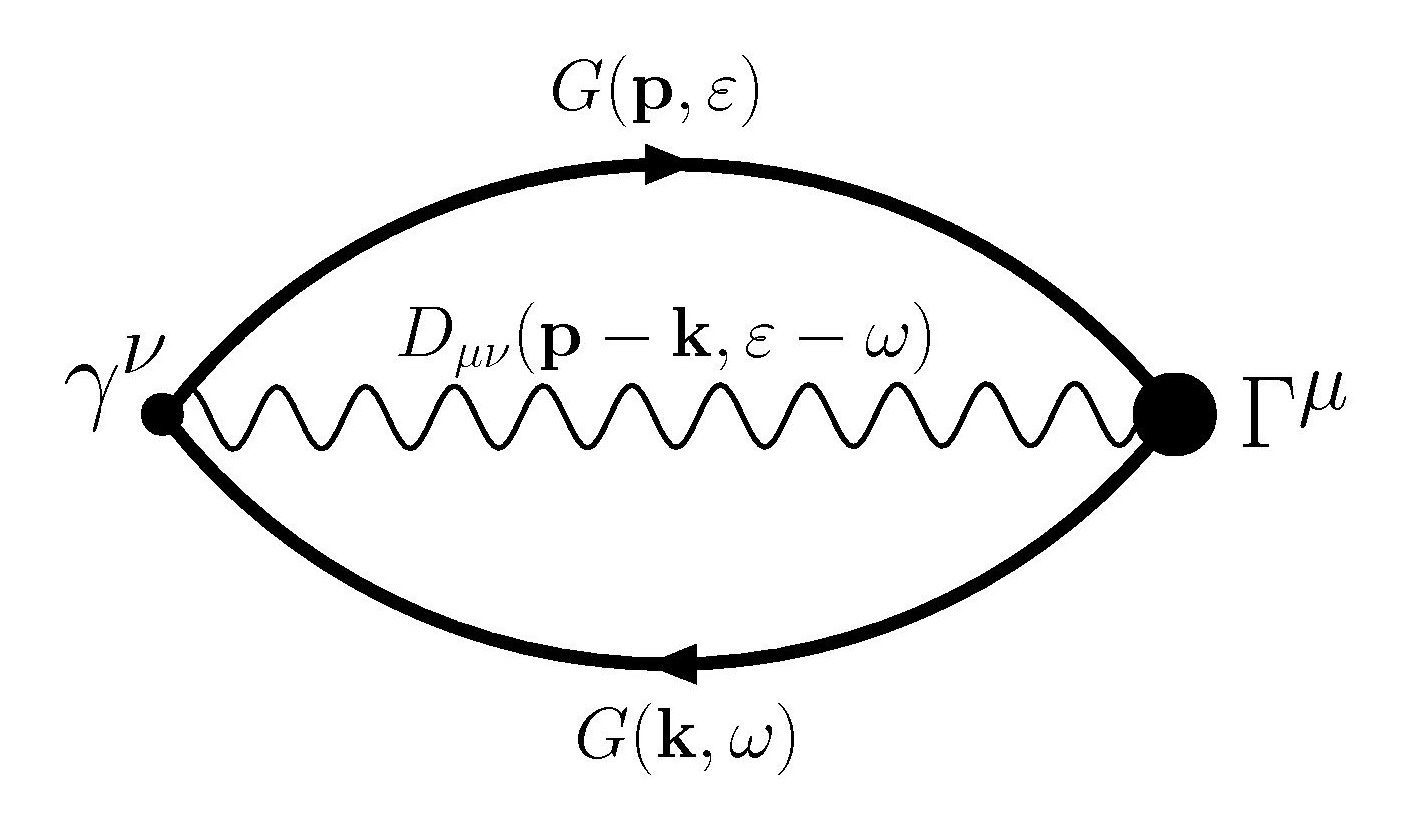}

\small{{\bf Fig. 1.} Exchange diagram.}

\vspace{0.1cm}

\hypertarget{fig2}{}
\includegraphics[scale=0.12]{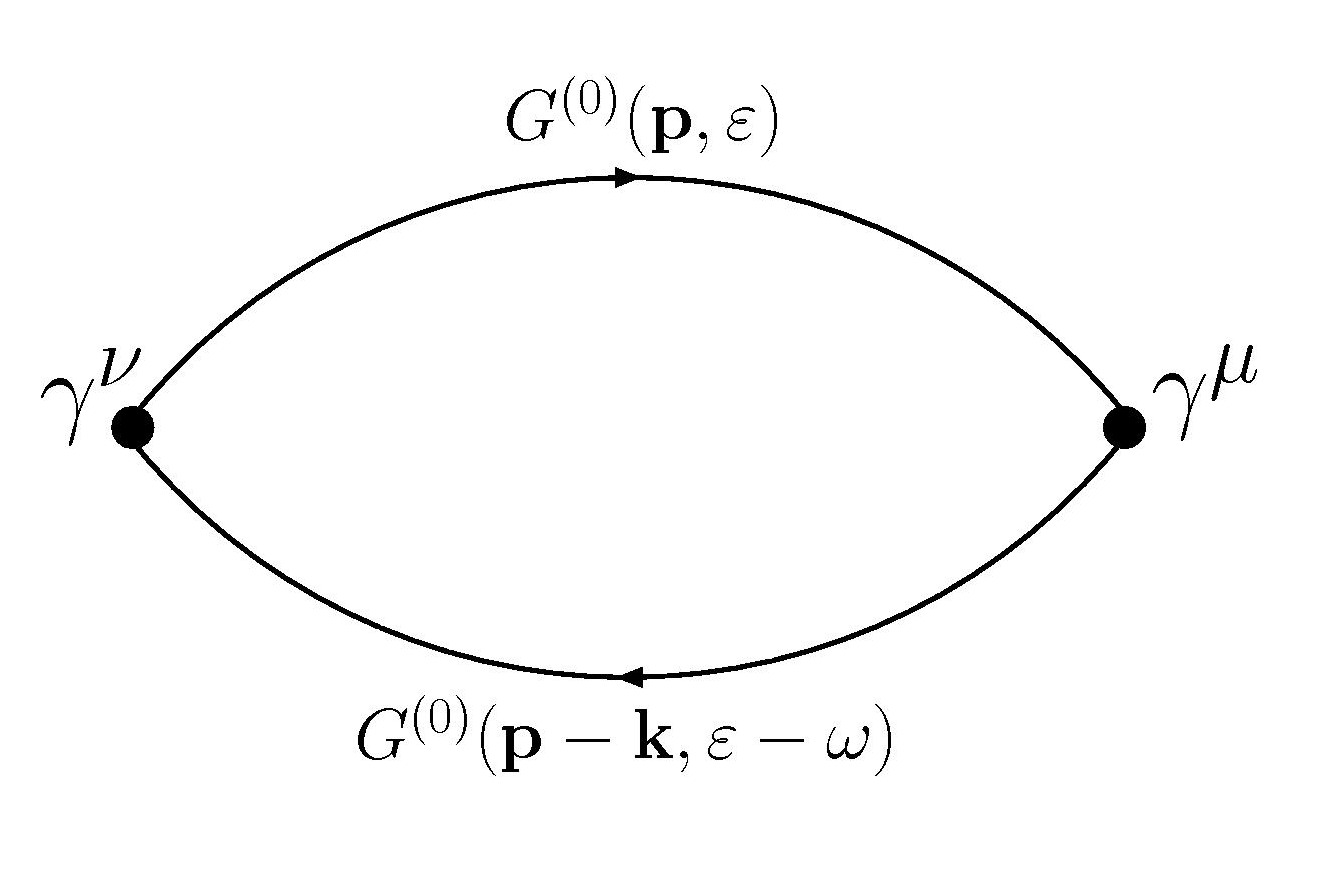}

\small{{\bf Fig. 2.} Zero-approximation polarization operator.}

\vspace{0.1cm}

\hypertarget{fig3}{}
\includegraphics[scale=0.12]{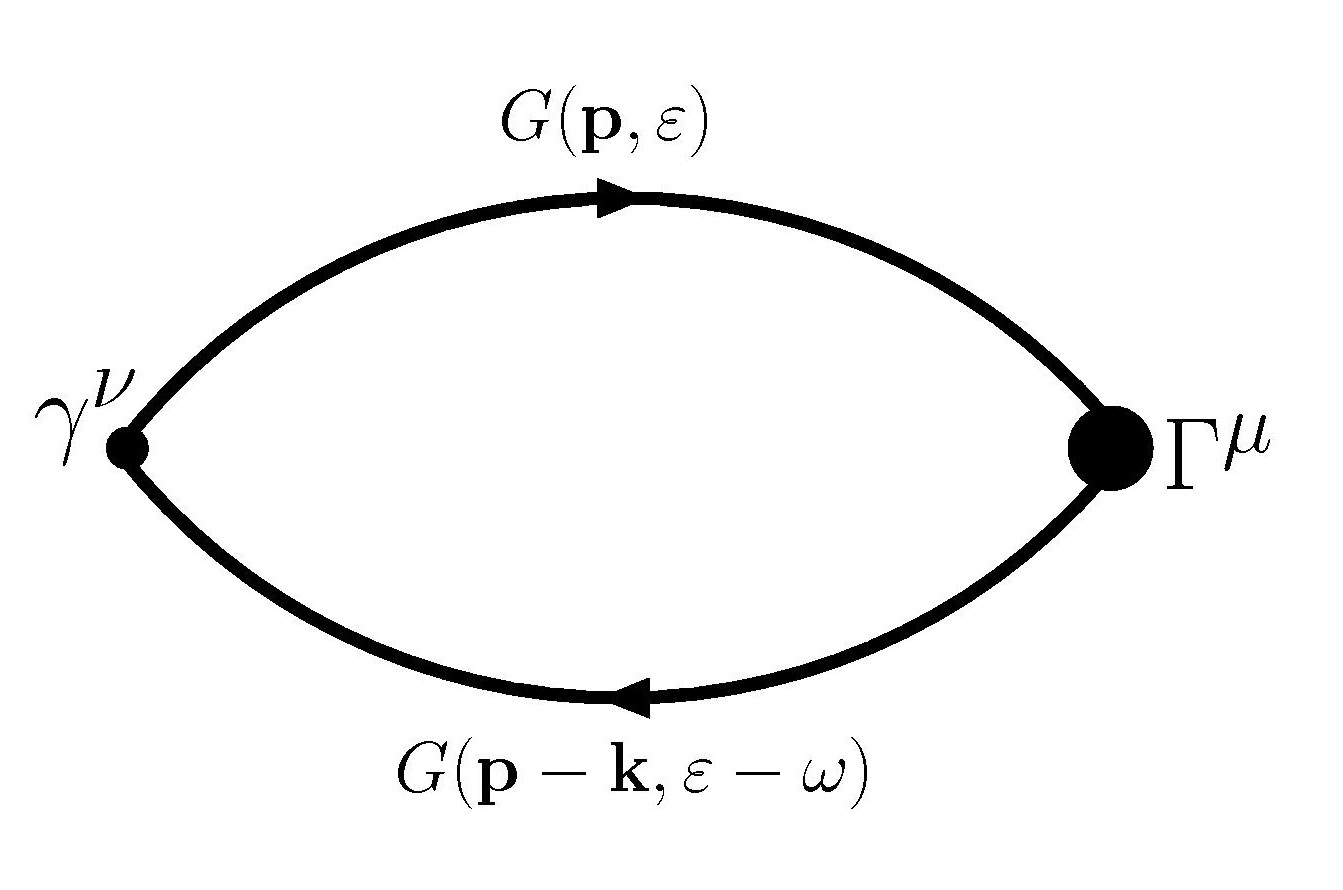}

\small{{\bf Fig. 3.} Total polarization operator.}
\end{center}
\vspace{0.5cm}
where $\widehat{p}=\gamma^\mu p_\mu, \ p_\mu=({\bf p},
i\varepsilon/u)$(pseudo-Euclidean metrics), $\delta_\pm=\delta
sign(\varepsilon_F\pm\varepsilon_{\bf p}), \ \delta\rightarrow+0$.
Evaluating the frequency integral \citep{Pechenik}
\begin{equation*}
\Pi^{(0)}_{44}\left({\bf k}, i\omega\right)=16\int\frac{d^2{\bf
p}}{(2\pi)^2}\frac{\theta\left(\left|{\bf
p}\right|-p_F\right)}{2\varepsilon_{\bf p}}
\end{equation*}
\begin{equation}\label{9}
\times\frac{\left({\bf k}\cdot{\bf p}\right)^2-\frac{\left|{\bf
k}\right|^2\varepsilon^2_{\bf p}}{u^2}} {\left(\left|{\bf
k}\right|^2+\frac{\omega^2}{u^2}\right)^2-4\left({\bf k}\cdot{\bf
p}-\frac{i\varepsilon_{\bf p}\omega}{u^2}\right)^2}.
\end{equation}
The total polarization operator $\Pi_{44}\left({\bf k},
i\omega\right)$ given by the diagram in \hyperlink{fig3}{Fig. 3} is
a renormalized quantity with respect to $\Pi^{(0)}_{44}({\bf k},
i\omega)$ due to the Coulomb interaction.

\begin{center}
3. RENORMALIZATION PROCEDURE
\end{center}

The relation for the constant $u=\frac{3}{2}\gamma a_0$ was obtained
analytically by linearizing the dispersion relation in the vicinity
of the $K$ and $K^\prime$ points of the Brillouin zone. The
dispersion relation was, in turn, found by solving the
Schr\"{o}dinger equation with the Hamiltonian in the tight-binding
approximation. However, the interaction between particles as the
interaction in a many-body system is disregarded in this
approximation. Thus, the constant $u$ must be renormalized taking
into account the Coulomb interaction. The renormalized value of u,
generally speaking, depends on the electron density.

\newpage

The problem is strictly formulated as a problem of solving the
system of integral equations for the exact Green’s function $G({\bf
p}, \varepsilon)$, exact photon propagator $D_{\mu\nu}({\bf q},
\Omega)$, and vertex function $\Gamma({\bf p}, \varepsilon; {\bf
p}^\prime, \varepsilon^\prime)$ (as well as for the polarization
operator and self-energy). The approximate solution of the equations
is possible, but seems too lengthy. We use here the renormalization
theory developed in QED. In fact, the exact Green’s function, photon
propagator, and vertex function are replaced by the respective
quantities in the lowest order in the interaction multiplied by the
renormalizing constants.

We proceed from two assumptions:

(i) the charge coincides with the physically observed charge
(ignoring the environment)
\begin{equation}\label{10}
e=e_0;
\end{equation}

(ii) the effective mass of particles remains equal to zero; i.e.,
the Coulomb interaction does not open a gap until the transition
point to the semimetallic state
\begin{equation}\label{11}
\Delta\equiv0.
\end{equation}

The gap opening is energetically unfavorable \citep{Ratnikov}, which
is confirmed in experiments.

Let us quote the known relations of the renormalization theory
\begin{equation}\label{12}
\Gamma^\mu=Z_1\gamma^\mu,
\end{equation}
\begin{equation}\label{13}
G=Z^{-1}_2G^{(0)},
\end{equation}
\begin{equation}\label{14}
D_{\mu\nu}=Z^{-1}_3D^{(0)}_{\mu\nu}.
\end{equation}
According to the renormalization theory, the charge satisfies the
relation \citep{Fradkin1}
\begin{equation}\label{15}
e=Z^{-1}_1Z_2Z^{1/2}_3e_0.
\end{equation}
Taking into account Eq. \eqref{10} and the Ward identity $Z_1=Z_2$,
we obtain $Z_3=1$, which was assumed in Eq. \eqref{6}. The Green’s
function satisfies the relation
\begin{equation}\label{16}
G=G^{(0)}+G^{(0)}\Sigma G
\end{equation}
with the formal solution
\begin{equation}\label{17}
G^{-1}=G^{(0)-1}-\Sigma,
\end{equation}
where $G^{(0)-1}=-u\widehat{p}$. Therefore, it should be expected
that the inclusion of the interaction results in the renormalization
of $u$, the only parameter in the dispersion relation, taking into
account Eq. \eqref{11}. Looking for the self-energy in the
form\footnote{Below, we show that $\Sigma$ is independent of the
frequency $\varepsilon$ and $\Sigma({\bf
p})=Au{\boldsymbol\gamma}{\bf p}$, which does not, however,
influence Eq. \eqref{19}, because $p_0$ at $p_4=ip_0$ should be
replaced by $\varepsilon/u^*$.}
\begin{equation}\label{18}
\Sigma({\bf p}, \varepsilon)=Au\widehat{p},
\end{equation}
we obtain
\begin{equation}\label{19}
G^{-1}({\bf p}, \varepsilon)=-u^*\widehat{p},
\end{equation}
where $u^*=(1+A)u$ is the renormalized $u$ value. The self-energy is
given by the expression (see \hyperlink{fig4}{Fig. 4})
\begin{equation*}
\Sigma({\bf p}, \varepsilon)=i\int\frac{d^2{\bf
q}d\Omega}{(2\pi)^3}\Gamma^\mu({\bf p}-{\bf q}, \varepsilon-\Omega;
{\bf p}, \varepsilon)G({\bf p}-{\bf q}, \varepsilon-\Omega)
\end{equation*}
\begin{equation}\label{20}
\times\gamma^\nu D_{\mu\nu}({\bf q}, \Omega).
\end{equation}
\begin{center}
\hypertarget{fig4}{}
\includegraphics[scale=0.15]{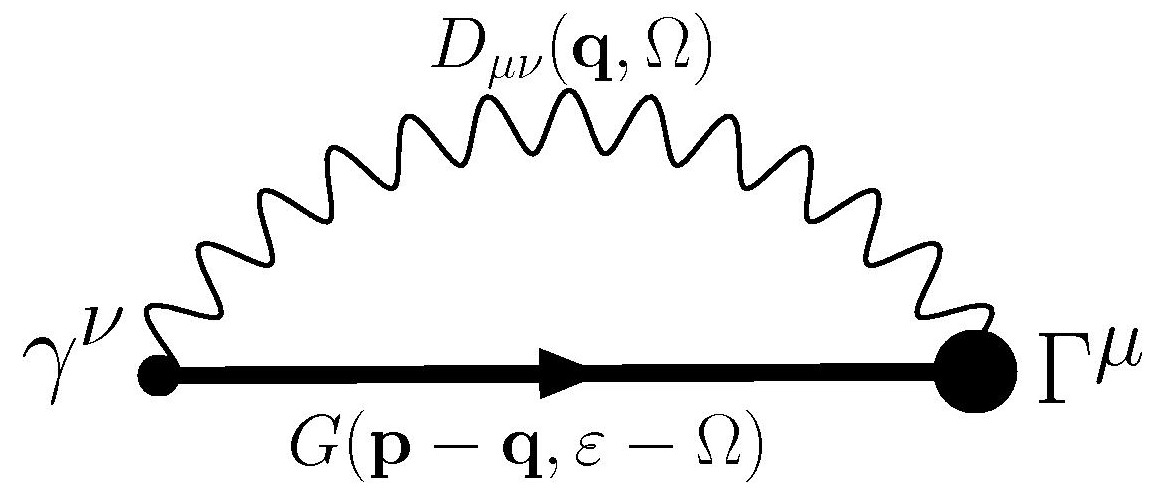}

\small{{\bf Fig. 4.} Self-energy.}
\end{center}
\vspace{0.5cm} Taking into account Eqs. \eqref{12} and \eqref{13}
and the Ward identity, we come to
\begin{equation}\label{21}
\Sigma({\bf p}, \varepsilon)=i\int\frac{d^2{\bf
q}d\Omega}{(2\pi)^3}\gamma^\mu G^{(0)}({\bf p}-{\bf q},
\varepsilon-\Omega)\gamma^\nu D^{(0)}_{\mu\nu}({\bf q}, \Omega).
\end{equation}
After simple calculations (see below), we obtain
\begin{equation}\label{22}
A=\alpha^*I\left(\frac{|\bf{p}|}{p_F}\right),
\end{equation}
\begin{equation}\label{23}
I(x)=\frac{1}{\pi}\int\limits_0^{\pi/2}\ln\left(\frac{1}{2}+\frac{1}{2}\sqrt{1-x^2\sin^2\varphi}\right)d\varphi,
\end{equation}
i.e., the renormalized dispersion relation has the form
\begin{equation}\label{24}
\varepsilon^*_{\bf p}=\pm u|{\bf
p}|\left(1+\alpha^*I\left(\frac{|\bf{p}|}{p_F}\right)\right).
\end{equation}

\begin{center}
\textit{Self-Energy Calculation}
\end{center}

Evaluating the pole integral with respect to $\Omega$ in Eq.
\eqref{21}, we obtain the expressions for the imaginary and real
part, which will be considered separately:
\begin{equation}\label{21a}
Re\Sigma({\bf p})=-\int\frac{d^2{\bf
q}}{(2\pi)^2}\frac{u\gamma^1(p_1-q_1)}{2\varepsilon_{\bf p-q}}V({\bf
q})\theta\left(|{\bf p-q}|-p_F\right),
\end{equation}
\begin{equation*}
Im\Sigma({\bf p})=-\int\frac{d^2{\bf
q}}{(2\pi)^2}\frac{u\gamma^2(p_2-q_2)}{2\varepsilon_{\bf p-q}}V({\bf
q})\theta\left(|{\bf p-q}|-p_F\right)
\end{equation*}
\begin{equation}\label{21b}
+\frac{1}{2}\gamma^0\int\frac{d^2{\bf q}}{(2\pi)^2}V({\bf
q})\theta\left(|{\bf p-q}|-p_F\right).
\end{equation}
Let ${\bf p}$ be directed along $q_x$, then $p_1=p$ and $p_2=0$.
Integral \eqref{21a} is the difference between two integrals (with
$p_1$ and $q_1$ in their numerators), both diverging at $q\gg p$.
Let us expand the integrand of the integral with $q_1$ in the series
to the $\sim p/q^2$ term, which gives
\begin{equation*}
\frac{\alpha^*}{4}u\gamma^1p\ln{\frac{q_c}{p_F}},
\end{equation*}
where the upper momentum cutoff $q_c\simeq2\pi/3\sqrt{3}a_0$ is
introduced, which is on the order of the half-distance between the
neighboring $K$ and $K^\prime$ points of the Brillouin zone (it is
the momentum at which the linear dispersion relation breaks down).

To evaluate the integral with $p_1$, let us introduce the
dimensionless variable
$z=\frac{q-p\cos{\varphi}}{p|\sin{\varphi}|}$. By taking the
integral with respect to $z$, the integral with respect to the angle
is reduced to the form
\begin{equation*}
-\frac{\alpha^*}{4\pi}u\gamma^1p\int\limits_0^{2\pi}
d\varphi\ln{\left(\frac{z_c+\sqrt{1+z^2_c}}{z_0+\sqrt{1+z^2_0}}\right)},
\end{equation*}
where
\begin{equation*}
z_0=\frac{\sqrt{p^2_F-p^2\sin^2{\varphi}}}{p|\sin{\varphi}|},
\hspace{0.5cm} z_c=\frac{q_c-p\cos{\varphi}}{p|\sin{\varphi}|}.
\end{equation*}
Taking into account that $z_c\gg1$, we may write
\begin{equation*}
-\frac{\alpha^*}{4\pi}u\gamma^1p\int\limits_0^{2\pi}
d\varphi\ln{\left(z_c+\sqrt{1+z^2_c}\right)}
\end{equation*}
\begin{equation*}
\approx-\frac{\alpha^*}{2}u\gamma^1p\ln{\frac{q_c}{p_F}}+\frac{\alpha^*}{4\pi}u\gamma^1p
\int\limits_0^{2\pi}d\varphi\ln{\frac{y|\sin{\varphi}|}{2}}.
\end{equation*}
Combining all these integrals together, we obtain the real part in
the form
\begin{equation*}
Re\Sigma({\bf p})=-\frac{\alpha^*}{4}u\gamma^1p\ln{\frac{q_c}{p_F}}+
\end{equation*}
\begin{equation}\label{21c}
+\frac{\alpha^*}{\pi}u\gamma^1p
\int\limits_0^{\pi/2}d\varphi\ln{\left(\frac{1}{2}+
\frac{1}{2}\sqrt{1-\left(\frac{p}{p_F}\right)^2\sin^2{\varphi}}\right)}.
\end{equation}
The first term in Eq. \eqref{21c} must be disregarded due to the
following reasons:

(i) $q_c\gg p_F$, and $\ln{\frac{q_c}{p_F}}$ may be arbitrarily
large at an arbitrarily small $p_F$; the renormalization coefficient
for u becomes negative and, thereby, the further calculation of the
ground-state energy $E_{gs}$ is senseless;

(ii) the corresponding contribution to the coefficient $A$ is
independent of the particle momentum; one may expect that the
particles with $p\rightarrow0$ are insensitive to the
renormalization of $u$ and the particles with $p\rightarrow p_F$ are
the most sensitive to this renormalization. Thus, the condition
$A(0)=0$ must be fulfilled.

The integration with respect to the angle in the first integral in
Eq. \eqref{21b} yields zero; the non-zero contribution is
\begin{equation}\label{21d}
Im\Sigma({\bf
p})=\frac{\alpha^*}{2}\gamma^0uq_c-\frac{\alpha^*}{\pi}\gamma^0E\left(\frac{p}{p_F}\right)up_F,
\end{equation}
where $E(x)$ is the complete elliptic integral of the second kind.
The first term is removed by the requirement $Im\Sigma_{Reg}(0)=0$,
i.e.,
\begin{equation*}
Im\Sigma_{Reg}({\bf
p})=\frac{\alpha^*}{\pi}\left[\frac{\pi}{2}-E\left(\frac{p}{p_F}\right)\right]\gamma^0up_F,
\end{equation*}
which represents the momentum-dependent shift of the frequency
$\varepsilon$ in the renormalized Green’s function
\begin{equation*}
G^{-1}({\bf p}, \varepsilon)=-(1+A)u{\boldsymbol\gamma}{\bf
p}-i\left\{\gamma^0\varepsilon+Im\Sigma_{Reg}({\bf p})\right\},
\end{equation*}
However, the shift is a slowly varying function of the momentum and,
therefore, may be replaced by its average value, i.e., by the
constant by which the integration with respect to the frequency may
be shifted when calculating the diagrams. Thus, the shift may be
disregarded and we arrive at the result given by Eqs. \eqref{22} and
\eqref{23}.

\newpage

\begin{center}
4. GROUND-STATE ENERGY
\end{center}

The average kinetic energy is now equal to
\begin{equation*}
E_{kin}=\frac{2}{3}up_F+
\end{equation*}
\begin{equation*}
+\frac{2\alpha^*}{\pi}up_F\int\limits_0^1x^2dx
\int\limits_0^{\pi/2}\ln\left(\frac{1}{2}+\frac{1}{2}\sqrt{1-x^2\sin^2\varphi}\right)d\varphi
\end{equation*}
\begin{equation}\label{25}
\approx\left[\frac{2}{3}-0.0342\alpha^*\right]up_F,
\end{equation}
Hence, the contribution of the renormalization is $\lesssim0.1$; the
respective contribution to the exchange and correlation energies is
expected to be of the same order of magnitude (or even smaller).
Therefore, to simplify further calculations, it reasonable to retain
the linear form of the dispersion relation by averaging Eq.
\eqref{22} in $|\bf{p}|$ and replacing $u^*$ with
$\overline{u}^*=(1+\overline{A})u$, where
\begin{equation*}
\overline{A}=\frac{\alpha^*}{\pi}\int\limits_0^1dx
\int\limits_0^{\pi/2}\ln\left(\frac{1}{2}+\frac{1}{2}\sqrt{1-x^2\sin^2\varphi}\right)d\varphi\approx
\end{equation*}
\begin{equation}\label{26}
\approx-0.0269\alpha^*.
\end{equation}
The renormalization of Eq. \eqref{6} for the exchange energy results
in the equality $E_{exch}=Z^{-1}_2E^{(0)}_{exch}$, where
$E^{(0)}_{exch}$ is the exchange energy calculated with the
nonrenormalized Green’s function, photon propagator, and vertex
function:
\begin{equation}\label{27}
E^{(0)}_{exch}=-\frac{\alpha^*J}{2\pi}up_F,
\end{equation}
where
\begin{equation*}
J=\int\limits_0^1dx\int\limits_0^1dy\int\limits_0^{2\pi}d\chi\frac{(1+\cos\chi)xy}{\sqrt{x^2+y^2-2xy\cos\chi}}=
\frac{8}{3}\left({\mathcal G}+\frac{1}{2}\right),
\end{equation*}
with ${\mathcal G}=0.915965\ldots$ being the Catalan’s constant.

The correlation energy may be calculated in the second order of the
perturbation theory. The corresponding diagrams are renormalized by
the factor $Z^{-2}_2$; thus,
$E_{corr}=Z^{-2}_2E^{(0)}_{corr}$.\footnote{The number of
renormalized vertices is half the number of renormalized Green’s
functions and the other half of the vertices remain nonrenormalized.
Thus, the nth order diagram is renormalized by a factor of
$Z^{-n}_2$.} The further calculation with the use of the asymptotic
expressions for the polarization operator in the lowest order in the
interaction at low and high transferred momenta leads to the
expression \citep{Ratnikov}
\begin{equation}\label{28}
E^{(0)}_{corr}=-\frac{\alpha^{*2}\nu}{128\pi}\left(\frac{3\pi}{8}-\frac{25}{27}-\frac{1}{27\nu}\right)up_F.
\end{equation}
The final result for the ground-state energy reads
\begin{equation*}
E_{gs}=\left[\sqrt{2\pi}\left(\frac{2}{3}-0.0342\alpha^*\right)-\frac{8\left({\mathcal
G}+\frac{1}{2}\right)}{3\sqrt{2\pi}}\right.
\end{equation*}
\begin{equation*}
\times\frac{\alpha^*}{1-0.0269\alpha^*}-\frac{\nu}{64\sqrt{2\pi}}\left(\frac{3\pi}{8}-
\frac{25}{27}-\frac{1}{27\nu}\right)
\end{equation*}
\begin{equation}\label{29}
\left.\times\frac{\alpha^{*2}}{(1-0.0269\alpha^*)^2}\right]u\left(\frac{n_{2D}}{\nu}\right)^{1/2}.
\end{equation}

\begin{center}
5. TRANSITION TO A SEMIMETALLIC STATE
\end{center}

It is seen from Eq. \eqref{29} that the coefficient of
$u\left(n_{2D}/\nu\right)^{1/2}$ changes its sign at
$\alpha^*_0\approx1.0204$ for $\nu=2$.\footnote{For $\nu=1$, we
obtain $\alpha^*_{01}\approx1.0214$ and $\alpha^*_{01}>\alpha^*_0$,
which is important if the transition is approached from the
$\alpha^*<1$ side. In this case, the transition from the
spin-unpolarized phase to the spin-polarized one occurs due to the
change in the sign of $E_{gs}$.} When this coefficient is negative,
the creation of electron– hole pairs in the system of the
two-dimensional massless Dirac fermions becomes favorable taking
into account that the absolute value of the ground-state energy is a
monotonically increasing function of the particle density. This
behavior is a manifestation of the system instability with respect
to the Coulomb interaction. In this case, a phase transition occurs.

As mentioned above, the gap opening is energetically unfavorable;
therefore, this is a transition from a zero-gap semiconductor state
to a semimetallic one. Remarkably, the transition occurs depending
on the $\alpha^*$ value, which in turn depends on the substrate
permittivity. Thus, the substrate material determines whether
graphene has semiconductor or semimetallic properties. It is also
seen that the spin-unpolarized state with $\nu=2$ is energetically
more favorable at a positive value of the ground-state energy,
whereas the spin-polarized state with $\nu=1$ is more favorable at
the opposite sign.\footnote{A spin factor of 2 is already taken into
account in the formula for $p_F$. Thus, in our notation, $\nu$
should be replaced by $\nu_{e,h}/2$ in the spin-polarized case
without changing the formulas.} As a result, the additional
transition to the spin-polarized state occurs simultaneously with
the transition to the semimetallic state.

The closeness of the parameter $\alpha^*_0$ to unity indicates that
there is a close analogy between the case under consideration and
the instability of the Coulomb field of the charge $Z=137$ in QED
with respect to the spontaneous creation of electron–positron pairs
(the fine-structure constant effectively approaches unity). It is
possible that the exact value of $\alpha^*$, at which the described
transition occurs, is actually equal to unity and the insignificant
deviation of $\alpha^*_0$ from unity obtained in this work is due to
the inaccuracy of the approximation used.

\begin{center}
6. POSSIBLE EXPERIMENTAL OBSERVATIONS
\end{center}

According to the experiments, the $\alpha^*$ value depends on the
substrate material. For example, $\kappa_{eff}=5$ and
$\alpha^*\approx0.44$ for the $SiO_2$ substrate and $\kappa_{eff}=3$
and $\alpha^*\approx0.73$ \citep{Iyengar} for the $SiC$ substrate,
but the condition $\alpha^*\simeq1$, which corresponds to
$\kappa_{eff}\simeq2$, is required for the transition.

Formally, $\kappa_{eff}=1$ in a vacuum and graphene must be a
semimetal. The overlap of the valence band and conduction band in
the semimetallic state may be estimated as \citep{Ratnikov}
\begin{equation}\label{30}
\delta E\simeq\left(b-\frac{1}{b}\right)up_F,
\end{equation}
where $b=\alpha^*\alpha^*_0$, which is only meaningful at
$\alpha^*>\alpha^*_0$. It follows from Eq. \eqref{30} that $\delta
E\propto n^{1/2}_{2D}$ ($p_F=\sqrt{2\pi n_{2D}}$). Electrons and
holes may appear as charge carriers in graphene at T = 0 when it is
situated on a substrate and the electric field is applied (the
electric field effect). In this case, the concentration of the
charge carriers is proportional to the applied voltage,
$n_{2D}\propto V_g$ \citep{Novoselov}. In the absence of a
substrate, $n_{2D}\equiv0$ and $\delta E\equiv0$ and graphene in a
vacuum remains a zero-gap semiconductor.\footnote{There is also a
purely technological difficulty in performing the experiment with
graphene suspended in a vacuum: the presence of a large number of
defects such as vacancies and local ruptures is possible. It is not
excluded that the graphene film is bent, creating a geometric
potential. If all of these difficulties were overcome, electrons
could be “sputtered” onto graphene, the necessity of the substrate
as a source of the charge carriers disappears, and the semimetallic
state of graphene becomes possible.} Thus, a substrate material with
a sufficiently small $\kappa_{eff}$ value should be found. It could
be, e.g., a substrate with regularly spaced pin holes or some
metamaterials. The application of an electric field is unnecessary
for the appearance of charge carriers in graphene in the case of the
second-type contact of graphene with the substrate, i.e., when the
$\varepsilon=0$ level of graphene does not lie in the band gap of
the substrate material.

\begin{center}
7. DISCUSSION
\end{center}

The main qualitative result of this work is the presence of the
phase transition from a zero-gap semiconductor to a semimetal in the
two-dimensional case at $\alpha^*\simeq1$. The zero-gap
semiconductor phase is stable at $\alpha^*\leq1$,\footnote{When
$\alpha^*$ exceeds unity by an infinitesimal value, the system
becomes unstable. However, at the exact equality $\alpha^*=1$,
$E_{gs}\equiv0$ should be expected and the system still remains
stable.} which is an analog of the stability criterion for the
system of three-dimensional non-relativistic fermions
\begin{equation*}
\frac{1}{\varepsilon_{tot}({\bf q}, 0)}\leq1,
\end{equation*}
where $\varepsilon_{tot}({\bf q}, 0)$ is the statistical limit of
the total dielectric function.

A semimetal–semiconductor transition in disordered degenerate
semiconductors was studied by Fradkin \citep{Fradkin}. The HgTe
(III–V semiconductors) and SnTe (IV–VI semiconductors) alloys, as
well as the twodimensional graphite, i.e., graphene, were considered
as the examples of such systems. The transition appears due to the
scattering of the charge carriers on a random potential and is
caused by the carrier localization, but the Coulomb and spin–orbit
interactions were disregarded in the model. In this work, the
influence of the disorder in graphene on its transport properties is
not considered, but it is shown that the inclusion of the Coulomb
interaction leads to a qualitatively similar result.

Abrikosov and Beneslavski\u{\i} \citep{Abrikosov} considered the
case of a Fermi point, a single-point touching of the conduction and
valence bands at $\varepsilon_F=0$. In this case, the Fermi energy
does not enter the Green’s function and the subsequent calculations
were performed at $p_F\equiv0$, which is essentially different from
the problem considered in this work, namely, graphene on a substrate
with a non-zero concentration of charge carriers, i.e., $p_F\neq0$.

\begin{center}
8. CONCLUSIONS
\end{center}

In this work, the ground-state energy of an electron (hole) gas was
calculated including the renormalization of the quantity $u$. The
stability of a zero-gap semiconductor phase with respect to the
transition to a semimetallic state was analyzed on the basis of the
calculation. The condition of the transition appearance was obtained
in the form $\alpha^*\simeq1$. Possible systems for the experimental
observation were pointed out.

I am deeply grateful to A.P. Silin for valuable advice during the
course of this work and fruitful discussions of the results.

\vspace{1cm}\hspace{4.2cm}\textit{Translated by A. Safonov}
\end{document}